\documentstyle[12pt,preprint,aps,floats]{revtex}
\tighten

\begin{document}

\preprint{
\noindent
\hfill
\begin{minipage}[t]{3in}
\begin{flushright}
TUM-HEP-278/97\\
RU--97--47\\
hep-ph/9707249\\
July 1997\\
\vspace*{2cm}
\end{flushright}
\end{minipage}
}

\draft

\title{Gravitational divergences 
as a mediator of \\ supersymmetry breaking}
\author{Hans Peter Nilles$^{a}$ and Nir Polonsky$^{b}$}
\address{${}^{a}$ Physics Department, Technische 
Universit$\ddot{a}$t M$\ddot{u}$nchen,
D-85748 Garching, Germany\\
and
Werner-Heisenberg-Institut,
Max-Planck-Institut f$\ddot{u}$r Physik, 
D-80805 M$\ddot{u}$nchen, Germany}
\address{${}^{b}$Department of Physics and Astronomy, 
Rutgers University, Piscataway, NJ 08855-0849, USA 
}

\maketitle

\begin{abstract}
Gravitational divergences associated 
with singlet fields in supersymmetric theories are reexamined,
and their possible contributions to the low-energy effective theory
are pointed out.
We demonstrate that
such divergences are not necessarily harmful and that
Planck-scale physics could play
an important role in  models of
low-energy supersymmetry breaking via a radiatively induced 
tadpole term  in the scalar potential. 
In this case, gravitational divergences play the role of 
the supersymmetry breaking mediator,
leading to a simple realization of
the so-called messenger  model.
We also point out
a new mechanism for the generation
of mass terms for the Higgs fields in models of low-energy
supersymmetry breaking, as well as 
a horizontal messenger mechanism  in which the horizontally
charged scalars are rendered heavy. 
Implications to the flavor problem in supersymmetric models
are also discussed.

\end{abstract}
\pacs{   }

Softly broken $N = 1$ supersymmetric theories
are only logarithmically divergent 
and can naturally accommodate weak-scale scalar fields.
In particular, supersymmetry
offers a well-defined framework for embedding and for extrapolating
to high energies the Standard Model of electroweak
and strong interactions (SM), which contains a  Higgs boson.

Supersymmetric extensions of the standard model 
postulate an equal number of fermionic and bosonic
degrees of freedom in the the low-energy theory, 
leading to the exact cancelation of 
quadratic divergences (which are not gravitational) in both
supersymmetric and softly broken theories.
The minimal supersymmetric extension of the standard model (MSSM)
essentially compliments each SM fermion (boson) with a boson 
(fermion) superpartner, and a plethora of new soft mass (and possibly
mixing and phase) parameters are needed to 
describe the superpartner potential. 
It must also contain  a pair of SU(2)
Higgs doublets with an opposite hypercharge 
(due to gauge anomalies involving the Higgs fermions),
as well as  the corresponding
supersymmetry conserving mass term, $\mu H_{1}H_{2}$, which renders
the Higgs fermions sufficiently massive.

The MSSM  provides a well-motivated and a well-defined framework 
for the study and search of physics beyond the
SM and has attracted great attention in recent years.
Nevertheless, the MSSM, or any other supersymmetric extension of the SM,
is most probably only the low-energy effective approximation
of a more fundamental theory which may explain the origin
of its many parameters.
Since it contains only logarithmic divergences,
it is rather insensitive to the details of the high-energy theory. 
This situation is a consequence of 
supersymmety, as explained above,
but also of the fact that the MSSM
does not contain a field which is a singlet under all global and 
local symmetries ({\em i.e.}, a universal singlet).
>From the model-building point of view, singlet fields -- whether fundamental or
composite objects -- are a useful
tool because of their trivial transformation properties,
and can be easily incorporated into
extended frameworks. 
Their introduction, however, reintroduces 
quadratic divergences to the low-energy theory, {\em i.e.}, 
quadratically divergent terms
associated with singlet fields
appear in the softly broken theory~\cite{PS,NSW,Lahanas,Ell,BP,Jain,BPR,Kras}.
The divergences are induced by the gravitational interactions
of the singlet.
(Similar divergences also appear in the vacuum energy~\cite{CKN,BPR}.)
The additional divergences render the singlet sensitive to the physics
at very high energies. This is similar to the situation
with a fundamental Higgs boson field in the SM (in the absence of
supersymmetry).

The singlet could become heavy, in which case, it either decouples and
is irrelevant and harmless or it destabilizes the MSSM parameters (if
it is coupled to other low-energy fields and renders those fields
heavy). Of course, in the latter case one has to require the absence
of such a harmful singlet.  On the other hand, a universal singlet may
be safely incorporated into the low-energy theory in certain
situations.  Furthermore, it may even be the agent that parameterizes
the high-energy physics in its low-energy effective approximation.  In
this paper we will attempt to examine this possibility.  We will
investigate whether such a situation exists and if so, its
implications for the low-energy effective theory.  We will do so by
examining the divergent scalar potential of the universal singlet
focusing on the possible role of the quadratically divergent terms in
the low-energy theory.  In particular, we will reexamine the relations
between the potential and the scale of supersymmetry breaking,
$M_{SUSY}$.

Supersymmetry is broken in some sector of the theory at a scale
$M_{SUSY}^{2} \equiv \sqrt{3}m_{3/2}M_{P}$,
where  $m_{3/2}$ and $M_{P} = M_{\mbox{\tiny Planck}}/\sqrt{8\pi}$  
are the gravitino and the reduced Planck
mass, respectively, and we impose hereafter 
the condition of a vanishing cosmological constant.
If supersymmetry breaking, parameterized here by $m_{3/2}$,
is mediated to the SM (observable) sector by the gravitational
interactions of the SM fields, then 
the gravitino mass also parameterizes the
(soft) mass scale of  the 
superpartners of the ordinary fermions and of the gauge and Higgs bosons
(the sparticles). It leads to the constraint
$m_{3/2} \simeq m_{\mbox{\tiny sparticle}} \simeq m_{\mbox{\tiny weak}}$, 
which fixes the scale of the gravitino mass.
The one-loop divergences of
the cosmological constant introduce a $\sim 100\%$ ambiguity in
$m_{3/2}$ and in related low-energy parameters~\cite{CKN}, but do
not affect the qualitative relation between the scales. 
If the only interactions between the supersymmetry breaking
sector and the observable sector are gravitational, as one often assumes,
then supersymmetry breaking is said to have  taken place in a hidden sector. 
Such models 
are sometimes classified as
models of  gravitationally mediated supersymmetry breaking. 
It is impossible to realize in a satisfactory fashion 
a light universal singlet in the above  framework.
Radiatively induced quadratic divergences, which are proportional to 
$m_{3/2}^{2}$, destabilize the singlet field~\cite{Lahanas,Ell,BP,Jain,BPR}.
It slides at the quantum level to large field values 
and decouples from the low-energy theory.
As explained above, the singlet could be either irrelevant or harmful
in this case.
The decoupling of the singlet is, however, a result
of the {\em ad hoc} relation $m_{3/2} \simeq m_{\mbox{\tiny weak}}$, and 
it is not a generic property of the theory.

Below, we will examine possible frameworks in which  the 
constraint  $m_{3/2} \simeq m_{\mbox{\tiny weak}}$ is relaxed, {\em i.e.},
we will require only $m_{3/2} \lesssim m_{\mbox{\tiny weak}}$. 
Indeed, we find that the
divergent singlet might not decouple
from the low-energy theory. Furthermore, it may provide
an important model-building tool.
The relation between the value of the singlet field and  the
supersymmetry breaking scale, which are correlated by $m_{3/2}$,
allows for new mechanisms of gravitational mediation of supersymmetry
breaking in which $m_{\mbox{\tiny weak}}$ is 
a complicated function of $m_{3/2}$.
The superpartners of
the ordinary fermion and of the gauge and Higgs bosons are rendered
massive by, {\em e.g.}, gauge loops rather than by their
gravitational interactions.
The gravitational interactions, which affect 
only  the singlet field directly, are only a trigger 
for the generation of the MSSM soft supersymmetry breaking mass parameters.
The presence of a light universal singlet field
in the theory may be difficult to 
understand from the effective field theory point
of view. By definition, 
the theory does not contain any local or global symmetry
which  would prevent ${\cal{O}}(M_{P})$ tree-level mass terms for the singlet.
However, string theory often contradicts this somewhat naive argument,
and could contain in its low-energy limit 
massless states that cannot be understood from
the low-energy point of view.
In fact, such states may provide a unique low-energy
signature of string theory.
Note also that the singlet may not be a fundamental object.
This would be the case,
for example, if it is a bilinear composite given by duality
transformations, in which case its
mass could correspond to a nonrenormalizable operator
in the more fundamental strongly coupled theory~\cite{duality}
and could be highly suppressed\footnote{However, in this case
one has to forbid mass terms in the strongly coupled theory which
would correspond to a tree-level tadpole term in the weakly coupled
theory. If the strongly coupled theory contains also fundamental
singlets, fundamental-composite  mixing could arise from 
the duality transformations of Yukawa operators.}.
Here, we will simply assume
that the theory contains a light universal singlet, $S$,  and examine
its consequences.

The radiative emergence of dimensionfull parameters for the singlet field
in the low-energy effective 
theory is forbidden  by supersymmetry nonrenormalization
theorems. Nevertheless, terms
proportional to the supersymmetry breaking
parameter $m_{3/2}$ are not forbidden by any mechanism 
and could still appear at low-energy.
Their gravitational origin is imprinted in their proportionality
to the gravitino mass.
Indeed, it has been shown 
that gravitationally induced tadpole terms for the 
singlet superfield typically appear at 
one~\cite{Lahanas,Ell,BP,Jain} or two ~\cite{BPR} loop order.
It is instructive to construct the divergent potential in 
a simple example, which can then be generalized.
The tadpole terms could arise, {\em e.g.}, due to Planck mass  suppressed 
trilinear terms in the Kahler potential,
\begin{equation}
K = \sum_{I}\left[1  + \frac{\alpha_{I}}{M_{P}}(S + S^{\dagger})\right]
\Phi_{I}\Phi^{\dagger}_{I} 
+ ...,
\label{K1}
\end{equation}
where the dots denote terms suppressed by higher orders of the Planck
mass and include, in principle, mixing between hidden and
observable fields (possibly leading to a non-trivial Kahler curvature).  
The summation is over the chiral superfields
$\Phi_{I}$, $S$ is a singlet chiral superfield and 
$\alpha_{I}$ are ${\cal{O}}(1)$ dimensionless couplings.
The leading nonrenormalizable operators given in eq.~(\ref{K1})
are allowed only if $S$ is a gauge and global symmetry (universal) singlet,
which we assume hereafter.

The low-energy Lagrangian contains contributions
from the Kahler potential $K$, the so-called $D$-terms,
\begin{equation}
{\cal {L}}_{D} 
= \int d^{2}\theta d^{2}\bar{\theta}EK,
\label{Ld}
\end{equation}
where $\theta$ is the Grassman variable and $E$ is the superspace
density $E = e^{K/M_{P}^{2}}$.
In particular, eq.~(\ref{Ld})
includes the trilinear vertices $\alpha_{I}ES\Phi_{I}\Phi_{I}^{\dagger}/M_{P}$.
A tadpole diagram, which is
allowed by all symmetries, appears at one loop,
\begin{equation}
{\cal {L}}_{D} \sim
\frac{N}{16\pi^{2}}
\frac{\Lambda^{2}}{M_{P}}\int{d^{2}\theta d^{2}\bar{\theta}E(S + S^{\dagger})},
\label{loop}
\end{equation}
where we set, for simplicity, all dimensionless couplings to unity,
$N$ counts the number of massless fields $Q_{I} \in \Phi_{I}$,
and $\Lambda$ is the
momentum cutoff scale for the divergent loop diagram.

The superspace density can be expanded~\cite{BPR}
\begin{equation}
e^{K/M_{P}^{2}} = 1 + \frac{1}{M_{P}^{2}}\left[
\theta^{2}K_{i}F^{i} + \bar{\theta}^{2} K_{i^{*}}F^{i^{*}}
+ \theta^{2}\bar{\theta}^{2}
(K_{ij^{*}} + \frac{K_{i}K_{j^{*}}}{M_{P}^{2}})
F^{i}F^{j^{*}}\right],
\label{eK}
\end{equation}
where $K_{i}$ ($K_{i*}$) denotes derivatives with respect to
supersymmetry breaking field
$Z^{I} = z^{i} +\theta^{2}F^{i} \in \Phi_{I}$ ($Z^{I*}$) 
and $F^{i} \sim M_{SUSY}^{2}$.
Substituting (\ref{eK}) into (\ref{loop}),
taking $\Lambda \sim M_{P}$, as is appropriate
in the case of gravitational interactions (\ref{Ld}), and  
using $F^{i} \sim M_{SUSY}^{2} \sim m_{3/2}M_{P}$,
one derives the contribution to the low-energy potential
\begin{equation}
|\Delta V| \sim \frac{N}{16\pi^{2}}  
\frac{1}{M_{P}}\left\{K_{i*}m_{3/2}M_{P}F_{s} 
+ (K_{ij^{*}} + \frac{K_{i}K_{j^{*}}}
{M_{P}^{2}})m_{3/2}^{2}M_{P}^{2}s + h.c.\right\},
\label{eK3}
\end{equation}
where $S = s + \theta^{2}F_{s}$.

One could distinguish two obvious limits:
$(1)$ Planckian values for supersymmetry breaking fields: $K_{i}  \sim M_{P}$;
$(2)$ No  Planckian values for supersymmetry breaking fields: 
$K_{i} < M_{SUSY} \ll M_{P}$.
Taking the appropriate limits in eq.~(\ref{eK3}) one has in case $(1)$ 
\begin{equation}
|\Delta V| \sim \frac{N}{16\pi^{2}}\left[
m_{3/2}^{2}M_{P}s \pm m_{3/2}M_{P}F_{s} + h.c.\right].
\label{deltaV1}
\end{equation}
In case $(2)$, only the tadpole term, $m_{3/2}^{2}M_{P}s \sim
[M_{SUSY}^{4}/M_{P}]s$, is present.
In either case, the low-energy potential could be written as
$\Delta V = M_{P}\sum_{n=1}^{3}C_{n}m_{3/2}^{n}q_{i}^{3-n}$.
In this form the mixing between the heavy and light sectors is seen
more clearly. It is sufficiently suppressed if $m_{3/2} \rightarrow 0$,
but  otherwise it could  destabilize the light scalars $q_{i}$.
Note also that each term in eq.~(\ref{deltaV1})
carries a degree of freedom associated with its phase. 
Our phase choice below will correspond to a non-trivial
global minimum of the potential, {\em i.e.}, 
we explicitly assume that the divergence is relevant
to the low-energy theory.
(This would be the case, for example, if the phase is determined dynamically.)

The elevation of the spurion derivation used above
to a supergravity derivation  implies new divergent diagrams due to 
(divergent) renormalization of the kinetic terms.
The two sets of one-loop diagrams cancel~\cite{Jain}, unless
the Kahler curvature is non trivial in the appropriate 
direction~\cite{Jain,BPR}.
However, the cancelation is accidental and, in general, is limited to
one loop~\cite{BPR}. More generally one has
\begin{equation}
|\Delta V |\sim \frac{N}{(16\pi^{2})^{n}}\left[
m_{3/2}^{2}M_{P}s \pm m_{3/2}M_{P}F_{s} + h.c.\right],
\label{deltaV2}
\end{equation}
where $n$ denotes the loop order of the first non-vanishing
divergent contribution to the potential, and similarly
in case $(2)$.  We typically expect $n \leq 2$.

Indeed, if the scalar potential 
$V \sim m_{3/2}^{2}[s^{2} - M_{P}s]$ then $s \sim M_{P}$
decouples  from the low-energy theory. 
(The singlet may be stabilized in intermediate scales if
the scalar potential contains a quartic term.)
In the limit $m_{3/2} \rightarrow 0$
one can neglect any soft contributions $\propto m_{3/2}\ll\sqrt{m_{3/2}M_{P}}$
to the low-energy potential other than eq.~(\ref{deltaV2}),
provided that the superpotential is not trivial.
The low-energy theory is given in this case by the
effective superpotential and by the divergent corrections.
Here, we will assume for simplicity the limit of case $(2)$, 
{\em i.e.}, $K_{i}/M_{P}\rightarrow 0$, and consider
a simple-minded  toy model.
The  toy model is given by the superpotential 
\begin{equation}
W = W(S) + \lambda SV\bar{V},
\label{sp1}
\end{equation}
where $V = v + \theta^{2}F_{v}$ and $\bar{V}$ are 
a vector-like pair of charged chiral
superfields, and we choose
\begin{equation}
W(S) = \frac{\kappa}{3} S^{3}.
\label{sp2}
\end{equation}
The scalar potential for $s$ is given by eqs.~(\ref{deltaV2}) and (\ref{sp2}),
\begin{equation}
V(s) = -\frac{M_{SUSY}^{4}}{M_{P}}s + 
\left|\frac{\partial W(S)}{\partial S} \right|^{2}
= -\frac{M_{SUSY}^{4}}{M_{P}}s + \kappa^{2}s^{4}.
\label{Vs}
\end{equation}
We omit hereafter the
dimensionless factor 
$N/(16\pi^{2})^{n} \sim {\cal{O}}(1 - 10^{-4})$ in eq.~(\ref{deltaV2}).
It is absorbed in $M_{SUSY}^{4}$ and corresponds to an order of
magnitude ambiguity in $M_{SUSY}$. 
In general, we expect modification of the quartic coupling at the
quantum level
due the field dependent masses of $S$, $V$ and $\bar{V}$
in the one-loop effective potential.
This is relevant only if $\kappa^{2} \lesssim 1/64\pi^{2}$.
More typically, however, one expects that all dimensionless couplings
are ${\cal{O}}(1)$.

In deriving eq.~(\ref{Vs}) we implicitly assumed 
that in the global minimum $v$, $\bar{v}$ and $F_{v}$, $F_{\bar{v}}$
vanish. This is indeed the case if the charge breaking 
direction $F_{s}^{*} = \lambda v\bar{v}
+ \kappa s^{2} = 0$ is only a local minimum of the model.
The corresponding constraint reads
\begin{equation}
2\lambda > \kappa,
\label{charge}
\end{equation}  
which is easily satisfied and which we will assume.

It is now straightforward to minimize the potential (\ref{Vs}),
\begin{equation}
s = \left(\frac{M_{SUSY}^{4}}{4\kappa^{2}M_{P}}\right)^{\frac{1}{3}}.
\label{svev}
\end{equation}
Note that the result for $s$ is only weakly sensitive to $\kappa$.
The order of magnitude ambiguity in $M_{SUSY}$ corresponds 
to a similar ambiguity in $s$. It could be compensated, if desired,
by an appropriate  choice of $\kappa$.
The inclusion of all ${\cal{O}}(m_{3/2})$ soft terms would shift
$s$ by ${\cal{O}}(m_{3/2})$. We will verify that the shifts are
negligible when discussing specific examples.

We also have $F_{s}/s = \kappa s$.
The nonvanishing $S$ components generate masses for the vector-like scalar pair
$(v,\bar{v})$,
\begin{equation}
M^{2}_{v\bar{v}} \sim 
\left(\begin{tabular}{c c}
$\lambda^{2}s^{2}$& $\lambda F_{s}$\\
$\lambda F_{s}^{*}$ & $\lambda^{2}s^{2}$
\end{tabular}\right).
\label{Mv}
\end{equation} 
The diagonal term is a supersymmetric mass term,
{\em i.e.}, the corresponding fermions have a similar Dirac mass term
$\lambda s \tilde{v}\tilde{\bar{v}}$.
Similarly, field dependent masses are induced for $s$ and for its fermion
partner $\tilde{s}$, and are given by replacing $\lambda$ with
$\kappa/3$ ($\kappa$) in the diagonal (off-diagonal) 
terms in eq.~(\ref{Mv}).

In the remainder of this paper we will discuss the application
of our toy model to various (phenomenological) schemes of
supersymmetry breaking. The most obvious framework that merits
examination is that of low-energy supersymmetry breaking:
$M_{SUSY}^{2} \ll m_{\mbox{\tiny weak}}M_{P}$ 
and $m_{3/2} \rightarrow 0$.
Typically, one assumes that if supersymmetry is broken at low energy 
then it is broken in 
a secluded (rather than hidden) sector.
The secluded sector communicates via new gauge interactions
with a messnger sector, which, in turn, communicates
via the ordinary gauge interactions with the observable sector.
The new gauge and  messenger Yukawa interactions 
mediate the supersymmety breaking 
to a (SM) singlet messenger $N = n + \theta^{2}F_{n}$,  which
parameterizes the supersymmetry breaking in the messenger sector.
The singlet $N$ interacts also with SM non-singlet messenger fields $M$ 
and $\bar{M}$.
The Yukawa interaction $\lambda N M \bar{M}$ communicates the supersymmetry
breaking to the messengers $M$ and $\bar{M}$ as in eq.~(\ref{Mv}).
In turn, the vector-like pair $M$ and $\bar{M}$, 
which transforms under the SM gauge group, 
communicates the supersymmetry breaking
to the ordinary MSSM fields via gauge loops.
The gauge loops commute with flavor and, thus, 
the spectrum is charge dependent  but flavor diagonal,
if one ensures that all other possible 
contributions to the soft spectrum are absent or are strongly suppressed.
Such models~\cite{GM} are often referred to as 
``gauge mediation of supersymmetry breaking'' or messenger models.
The sparticle spectrum, and hence, the weak scale, are given
in this framework
by $m_{\mbox{\tiny weak}}  \sim m_{\mbox{\tiny sparticle}} 
\sim (\alpha_{i}/4\pi)(F_{n}/n)$,
where $\alpha_{i}$ is the relevant gauge coupling at the scale
$\Lambda_{N} = F_{n}/n \sim 10^{5}$ GeV. One also assumes a similar
scale for the supersymmetry breaking in the secluded sector,
$M_{SUSY} \sim \Lambda_{n}$ (but $M_{SUSY}$ could be one or two
orders of magnitude higher).

The universal singlet $S$ could couple in this framework 
to either the messenger fields,
$V\bar{V}  \rightarrow M\bar{M}$, or to the MSSM Higgs doublets,
$V\bar{V}  \rightarrow H_{1}H_{2}$. By properly assigning, {\em e.g.}, 
Peccei-Quinn or $Z_{3}$ charges, one could forbid one of the coupling
while allowing the other. Alternatively, one could assume that
one of the Yukawa terms is forbidden by the selection rules in a
string theory (see discussion above and Ref.~\cite{Kobayash}).

In the case $W = SH_{1}H_{2} + \frac{\kappa}{3} S^{3}$, the mass term
(\ref{Mv}) corresponds to the supersymmetry 
preserving  (usually denoted by $\mu$) and breaking  
(usually denoted by $m_{12}^{2}$, $m_{3}^{2}$, or $B$) 
mass terms in the MSSM Higgs potential. 
Substituting $M_{SUSY} \sim 10^{6}$ GeV in eqs.~(\ref{svev}) and
(\ref{Mv}) one has $\mu \sim m_{12} \sim m_{\mbox{\tiny weak}}$,
as required. 
In comparison, if $\mu$ and $m_{12}$
are generated  by (Yukawa) loops, which would be the most natural mechanism
in this framework,
they would typically be generated  at the same loop order and one has
$m_{12}^{2}/\mu \sim \Lambda_{n}$, which is
phenomenologically unacceptable (but see Ref.~\cite{Dine}). 
The universal singlet mechanism
offers a simple resolution of this situation.
(For other proposals, see Ref.~\cite{DGP}.)

Alternatively, one could couple the universal singlet
to the messenger fields, mediating supersymmetry
breaking to the messenger sector via the gravitationally
induced tadpole term in the scalar potential.
Gravitational divergences play now the role of 
the supersymmetry breaking mediator,
leading to a simple realization of the  a messenger  model.
One need not introduce any new gauge interactions
between the secluded  and messenger sectors in order
to mediate supersymmetry breaking to the messenger sector.
The secluded sector is promoted to a truly hidden sector
while the messenger sector, which now
has a minimal content $\{ S = N, V = M, \bar{V} = \bar{M}\}$,
can be embedded in the observable sector. This significantly
simplifies the theory and avoids potentially  dangerous
directions in field space~\cite{color}.
(Such directions could appear if there are additional Yukawa
interactions which are introduced in order to communicate  the
supersymmetry breaking from 
the subset of messengers which transform under the new gauge interactions
to the singlet $N$.)
Requiring $\Lambda_{s} = F_{s}/s \sim 10^{5}$ GeV we find\footnote{In the limit
$(1)$, {\em i.e.}, $K_{i}/M_{P} \sim 1$, we find $M_{SUSY} \sim 10^{5}$ GeV.}
$M_{SUSY} \sim 10^{8}$ GeV. The gravitino mass\footnote{
The MeV gravitino may significantly constrain the cosmology~\cite{cosmology},
however, we do not discuss the cosmological implications in this paper.} 
is in the MeV range, and all terms proportional to 
$m_{3/2}^{n}$ (aside from the tadpole) can be safely neglected.
We did not need to make any assumptions regarding the nature of supersymmetry
breaking in the hidden sector, and we did not alter the 
generic phenomenology of the models.
The embedding of only the messenger fields
in the SM sector offers a simple and essentially 
model-independent  framework in comparison to  interesting but
complicated proposals made recently
for  embedding both the SM and messenger sectors
in the supersymmetry breaking sector~\cite{embed}.

If one raises $M _{SUSY}$, there is a 
possible interplay between $m_{3/2}$ and 
$(\alpha /4\pi)\Lambda_{s}$ mass terms, which allows us to 
consider a new class of messenger models.
The messenger fields in this framework, $T$ and $\bar{T}$, 
are charged  only under a gauged horizontal symmetry and are SM singlets.
For concreteness, we will consider an SU(2) horizontal
symmetry under which the third family and Higgs fields are singlets.
The first and second family fields transform as 
doublets $(U_{1},\,U_{2})$, {\em etc.}, under the symmetry.
For $M_{SUSY} \sim 10^{10-11}$ GeV one has $\Lambda_{s} = F_{s}/s \gtrsim
10^{7}$ GeV, leading to multi-TeV masses for all horizontally charged
MSSM scalars (assuming that the horizontal gauge coupling
$\alpha_{H} \lesssim \alpha_{\mbox{\tiny weak}}$). As  a result,
generic MSSM contributions to flavor changing neutral currents
involving the first and
second family fermions ({\em i.e.}, the flavor problem)
are sufficiently suppressed, and there is no need to impose
mass universality (as in the previous case)
or fermion-sfermion mass alignment.
The scalar components of the Higgs and third family fields
are singlets  of the symmetry and are lighter with only gravitational 
masses $\sim m_{3/2} \sim m_{\mbox{\tiny{weak}}}$.
Thus, one avoids excessive fine-tuning in the Higgs potential 
(which radiatively couples to the third 
family fields through their Yukawa interactions).
The spectrum\footnote{We do not specify the mechanism for the
breaking of the horizontal symmetry. It could involve the $T$ fields if
the constraint (\ref{charge}) is not satisfied.
In a given model one would have to ensure that the horizontal scale
is sufficiently high to suppress contributions to flavor violation from
the horizontal sector.}
is described in Table \ref{table:spectrum}.
Although $m_{3/2}$ is not negligible in this case,
the corresponding shift in the intermediate-scale $s$ is negligible.
The horizontal messenger model offers a simple realization
of a model with heavy first two families of sfermions~\cite{effectivesusy},
which is motivated by the resulting decoupling of the superpartners
from sensitive low-energy flavor (and CP) observables.

\begin{table}
\caption{Mass scales in the horizontal messenger model.}
\label{table:spectrum}
\begin{tabular}{ll}
Particle & Mass scale\\
\hline
First and second family sfermions &
$\frac{\alpha_{H}}{4\pi}\Lambda_{s}$ \\
Third family sfermions & $m_{3/2}$ \\
Higgs fields & $m_{3/2}$ \\
SM gauginos& $m_{3/2}$ \\
$S,\,T,\,\bar{T}$ & $\Lambda_{s}$ \\
Horizontal gauge fields &  $\lesssim \Lambda_{s}$ \\
\end{tabular}
\end{table}

All mechanisms described above lead to simple realizations
of theoretical frameworks\footnote{Note that such
frameworks often involve very heavy sparticles, probes of which
were recently discussed in Ref.~\cite{CFP}.}
which attempt to solve the flavor problem in supersymmetric models
by either low-energy sfermion mass universality (gauge-mediation framework)
or decoupling (heavy sfermions - light Higgs framework).
Indeed, {\em a priory} there is no reason to expect sfermion mass
universality to hold.
In the horizontal messenger  framework the universality
assumption can be significantly relaxed (as in the models of
Ref.~\cite{effectivesusy}).
In the gauge-mediation framework universality is achieved
by ensuring the exclusiveness of the gauge-loop contributions
to the soft spectrum. A similar situation arises
in the gravitational mediation framework 
if the sfermion spectrum is determined primarily by gaugino loops. 
Nevertheless, generic non-universalities in the scalar spectrum might appear
in the gravitational mediation framework, in particular, if this framework 
is the low-energy limit of a string  theory~\cite{jan}.
Even if universality, {\em i.e.}, $K_{q_{i}q_{j}^{*}} = \delta_{ij}$,
is imposed by hand, it could be spoiled by
either large radiative corrections ({\em e.g.}, if the model  is
embedded in a grand unified theory~\cite{PP}), or more generically,
by the tower of nonrenormalizable operators in the Kahler potential
which mix heavy and light fields~\cite{SW} (recall that supergravity is
a nonrenormalizable theory). 
Attempts to impose universality in (perturbative) string theory
by assuming the dominance of supersymmetry 
breaking in  the universal dilaton field~\cite{jan}
are also subject to radiative corrections~\cite{jan2}. 
(In addition, no convincing  example of such models exist.)
Clearly, universality may be realized if the coupling of the various
scalar fields to the Goldstino component of the gravitino is dictated
by a symmetry. In string theory this would be the case if the gauge
quantum numbers are linked to the conformal weights, in which case,
the moduli contributions to the soft spectrum are charge dependent
but flavor diagonal, as in the gauge-mediation framework.

Here, we would like to show
that even if heavy-light mixing is eliminated order by order in $M_{P}^{-1}$,
radiatively generated mixing among the light fields themselves
could still spoil the universality assumption. 
The singlet field is replaced by
a  gauge invariant composite operator of the light fields,
{\em e.g.},  $QQ^{\dagger}/M_{P}$ .
The Kahler potential (\ref{K1}) is rewritten in this case as 
\begin{equation}
K = \sum_{I}\left[1  + \frac{\alpha_{I}}{M_{P}^{2}}QQ^{\dagger}\right]
\Phi_{I}\Phi^{\dagger}_{I} 
+ ...,
\label{K2}
\end{equation}
and includes mixing among  the light fields $\{Q_{I}\}$.
The $D$-terms (\ref{Ld}) now contain the soft mass parameters 
\begin{equation}
\Delta_{Q}^{2} \sim  \sum_{I}\frac{\alpha_{I}}{(16\pi)^{n}}
m_{3/2}^{2},
\label{delta}
\end{equation}
where the summation is over the light fields.
Note that flavor indices in eq.~(\ref{delta}) need not be diagonal,
leading to both non-universalities and explicit flavor off-diagonal masses.
For $m_{Q}^{2} \sim m_{3/2}^{2}$, the size of the effect depends
on the size of the counting parameter
$N = \sum_{I} \alpha_{I}$ and on the loop-order, {\em i.e.},
$\Delta_{Q}^{2}/m_{Q}^{2} \sim N/(16\pi)^{n}$. 
The effect could be significant in comparison
to the (rough) phenomenological upper bound of 
$\Delta_{Q}^{2}/m_{Q}^{2} \lesssim 10^{-3}$~\cite{pokorski}.

Some comments are in place.
Indeed, one expects the emergence of radiative corrections
if certain terms which are allowed by all symmetries are eliminated
by hand from the Kahler potential. (For example, this is the case 
of the universal singlet tadpole.) 
However, here we have a  somewhat special case.
The terms $ Q_{1}Q_{1}^{\dagger}Q_{2}Q_{2}^{\dagger}$
cannot be forbidden by any typical field-theory symmetry
and hence, there are typically divergent radiative corrections 
to $Q_{1}Q_{1}^{\dagger}$.
The above terms could presumably be absent in a string theory framework
in the case of T-duality.
Otherwise, the radiatively induced tadpole (for the bilinears)
aggravates the flavor problem in models
of gravity mediated supersymmetry breaking.
On the other hand, we saw above that a universal singlet 
tadpole term could offer
alternative frameworks in which gravity is only a trigger
for the generation of the MSSM parameters, and in which
the flavor problem is solved by low-energy universality or decoupling.

We note in passing that
the Kahler potential (\ref{K2}) also provides
a new mechanism for the  generation of the  Higgs mixing parameter
$m_{12}^{2}$. The bilinear 
$QQ^{\dagger}$ is substituted in this case 
with the gauge invariant bilinear $H_{1}H_{2}$.
If lepton number is not a symmetry of the Kahler potential,
the $LH_{1}^{\dagger}$ and $LH_{2}$ bilinears 
($L$ denotes a lepton doublet) would similarly generate
Higgs-lepton mixing in the scalar potential.

It should be noted that a given theory contains only one scale
for all tadpole operators which is given by $M_{SUSY}$. Only one of the
above mechanisms could be realized in a given model.
For example, our realization of the (ordinary) messenger model
cannot benefit from our proposal for the generation
of the Higgs mass parameters in models of low-energy supersymmetry
breaking. However, a new source of tadpole diagrams could
arise if there is a superpotential mixing 
between heavy and light singlets~\cite{PS,NSW,Lahanas}
(which are not necessarily universal singlets).
The scale of the tadpole in this case depends on the 
the heavy singlet mass $M_{H}$ rather than on $M_{P}$.
One could couple the light singlet $S_{l}$ 
to  a vector-like multiplet by extending  the
superpotential 
\begin{equation}
W = M_{H}S_{h}^{2} + m_{l}S_{l}^{2} + \lambda_{1}S_{h}^{3}
+ \lambda_{2}S_{h}^{2}S_{l} + \lambda_{3}S_{l}^{3} 
+ \lambda_{4}S_{h}S_{l}^{2}
\label{NSW}
\end{equation}
of Ref.~\cite{NSW}  with the appropriate
Yukawa term $W \rightarrow W + \lambda_{5} S_{l} V \bar{V}$.
One could construct, in principle, 
models with a few singlet fields and with both gravitational
and heavy singlet tadpoles which are effectively described by
\begin{equation}
|\Delta V |\sim 
m_{3/2}^{2}M_{P}s \pm m_{3/2}M_{H}F_{s_{l}} + h.c.
\label{deltaV3}
\end{equation}
(Note that the assumptions in Ref.~\cite{NSW}
correspond to our limit $(2)$ above.)
The singlet $s$ is as before and 
its value is as in eq.~(\ref{svev}), 
while $s_{l} \sim \sqrt{m_{3/2}M_{H}}$. 
In this hybrid scenario for the messenger models,
$S$ couples to $H_{1}H_{2}$ while $S_{l}$ couples to $M\bar{M}$.

In conclusion, gravitationally induced tadpole
terms in the low-energy theory aggravate the flavor
problem if 
$m_{\mbox{\tiny sparticle}} \simeq m_{3/2} $,
but may be  desirable if 
$m_{3/2} \ll m_{\mbox{\tiny sparticle}}$.
In the latter case the universal singlet does not decouple 
and it parameterizes the effects of
the heavy sector in the effective low-energy theory.
In specific examples we were able to realize:
\begin{enumerate}
\item
Mass terms for the Higgs fields if $M_{SUSY} \sim 10^{6}$ GeV;
\item 
Gravitational triggering of a gauge mediation framework if
$M_{SUSY} \sim 10^{8}$ GeV;
\item
A horizontal messenger model if $M_{SUSY} \sim 10^{10}$ GeV.
\end{enumerate}
All of the above mechanisms stem from a simple-minded toy model
and are characterized by their simplicity. 
They merit further investigation and 
one may attempt to incorporate them
individually (or simultaneously, if including heavy-light mixing)
into more complete models.

\acknowledgements
It is pleasure to thank Erich Poppitz for his comments and insight.
This work was supported 
by the European  Commission programs ERBFMRX-CT96-0045 and CT96-0090, 
a grant from Deutsche Forschungsgemeinschaft SFB-375-95,
and by the US National Science Foundation under grant NSF-PHY-94-23002.

\end{document}